\documentstyle[12pt]{article}
\begin{document}
\global\arraycolsep=2pt 
%
\thispagestyle{empty} 
\begin{titlepage}    
\topskip 4cm
\begin{center}
  {\Large{\bf 
   How parametric resonance  mechanism  follows quench mechanism
   in  disoriented chiral condensate}  }
\end{center}                                   
          
\vspace{0.8cm}
              
\begin{center}
Shinji Maedan            
   \footnote{ E-mail: maedan@tokyo-ct.ac.jp}   
           \\
\vspace{0.8cm}
{\sl  Department of Physics, Tokyo National College of Technology,
        Kunugida-machi, Hachioji, Tokyo 193-0997, Japan
                                   }
\end{center}                                               
            
\vspace{0.5cm}
              
\begin{abstract}
\noindent       
We show how parametric resonance mechanism follows quench mechanism in the classical linear sigma model.
The parametric resonance amplifies long wavelength modes of the pion for more than $ 10 ~{\rm fm/c} $.
The shifting from the quench mechanism to the parametric resonance mechanism is described by a time dependent
   quantity.
After the quench mechanism is over, that quantity has an oscillating part, which causes
   the parametric resonance.
Since its frequency is $ 2 \, m_\pi ~( m_\pi $ : pion mass ), very long wavelength modes such as
   $ k = 40 {\rm MeV} $ of the pion are amplified by the parametric resonance.
\end{abstract} 
\vskip 2.5cm
\begin{center}
           Quenching, Isospin
\end{center}
\vfill            
\end{titlepage}
%
%
\baselineskip=0.7cm
%
%
\setcounter{page}{1}
\section{Introduction}
The notion of chiral symmetry breaking has a significant meaning to comprehend hadron physics.
The lattice numerical simulations or theoretical studies predict that the chiral symmetry is restored
   under the extreme temperature.
One of the purposes of experiments at Relativistic Heavy Ion Collider (RHIC) is to produce the
   chiral symmetric phase.
With regard to those experiments, the idea of chirally misaligned vacuum, the disoriented chiral condensate
   (DCC)  \cite{rf:AnsRys,rf:BlaKrz,rf:Bjo,rf:KowTay,rf:RajWil395}, has been discussed.
   
Rajagopal and Wilczek carried out numerical simulation in the classical $ O(4) $ linear sigma model with 
  fields
    $ \phi  = (  \sigma , {\vec \pi} ) $, which is often used to study the dynamics of
    chiral condensate  \cite{rf:RajWil577}.
They show that, after quenching, long wavelength modes of the pion fields are amplified.
This strongly suggests the formation of large DCC domains.
Indeed, Asakawa, Huang and Wang verified later by numerical simulation that the DCC domains with
   $ 4 \sim5 ~{\rm fm} $ in size can be formed
   through a quench \cite{rf:AsaHuaWang}.
If such large domains are actually produced in experiments, the existence of DCC would be confirmed
   by measuring, for example, a pion emission ratio
   $ R \equiv n_{\pi^0} / ( n_{\pi^0}+n_{\pi^+}+n_{\pi^-}  )$ \cite{rf:AnsRys,rf:BlaKrz}.

Therefore, it is crucial for the formation of large DCC domains that long wavelength modes 
   of the pion are amplified after quenching.
According to Ref.\cite{rf:RajWil577}, one can understand the enhancement of such modes by considering the
following
   mechanism.
After quenching, the fields $ \phi $ roll down randomly in isospin space from the top of the Mexican hat
   potential.
During the rolling down, the long wavelength modes of the pion field have negative mass square 
    $ m^2_{\rm eff} < 0 $, so that these modes grow exponentially.
This is the mechanism Rajagopal and Wilczek argued, which we call quench mechanism.
 
However, the numerical simulation in Ref.\cite{rf:RajWil577} shows that long wavelength modes 
   of the pion are amplified continuously for more than $ 10~{\rm fm/c} \;$ even when the effective mass 
   becomes $ m^2_{\rm eff} > 0 $ after the rolling down.
How can we understand such a phenomenon?
It is necessary to answer this question since the formation of DCC domain is due to the enhancement of
   those modes.

There are some attempts to explain such a phenomenon, which say that the oscillation of the sigma field around
   the true vacuum causes the enhancement of the pion modes through parametric resonance
   \cite{rf:MroMul,rf:HirMin,rf:Kai}.
In these proposals, it is assumed that the field $\phi$ oscillates along the $\sigma$ direction
   around the minimum of its potential after the rolling down.
It is, however, not obvious that the direction of the $\phi$'s oscillation is $\sigma$, so that
   we do not adopt the assumption here.
   
In this letter, we make a investigation of the cause of the enhancement of pion's long wavelength modes 
   after the isotropic rolling down.
The classical $O(4)$ linear sigma model is also used in our study, where we do not suppose that the field
   $\phi$ oscillates along the $\sigma$ direction after the isotropic rolling down, as mentioned above.
The point is that how the long wavelength modes of the pion are continuously amplified following the quench
   mechanism.
In order to argue the quench mechanism, Rajagopal and Wilczek have analyzed the Euler-Lagrange equation for the
   pion modes with the approximation where the field  $ \phi^2 (x) $ is replaced by its spatial average
   $ \langle \phi^2 \rangle (t) $.
(In Ref.\cite{rf:RajWil577}, the spatial average $ \langle \phi^2 \rangle (t) $ has been calculated by numerical
   simulation.)
We also use the Euler-Lagrange equation with the same approximation to find how the pion modes develop after
   the isotropic rolling down.
In our approach, the issue is the time dependence of the $\phi$'s spatial average $ \langle \phi^2 \rangle (t) $,
   not the sigma field's one $ \langle \sigma \rangle (t) $ \cite{rf:MroMul,rf:HirMin,rf:Kai}.
The time evolution of the pion modes thus derived will be compared with the result of numerical simulation
   obtained by Rajagopal and Wilczek.
   
In the next section, after reviewing the quench mechanism argued by Rajagopal and Wilczek  \cite{rf:RajWil577},
   we describe how the parametric resonance mechanism follows the quench mechanism.
The time evolution of the zero mode of the pion will be discussed in Section 3.
The last section is devoted to summary and discussion.
%
%
%
%
%
\section{Parametric resonance  after isotropic
            rolling down of fields}
The linear sigma model is
\begin{equation}
 {\cal L}= {1\over2} \partial^\mu \phi \partial_\mu \phi -{1\over4} \lambda
  (\phi ^2 -v^2)^2 + H \sigma,
  \label{ba}
\end{equation}
where $\phi = ( \sigma \, , \vec \pi \,)$.
Since an explicit chiral symmetry breaking term $H \sigma$ exists, the pion has mass
   $ m_\pi^2 = H / f_\pi $ and the true vacuum is $\phi= ( f_\pi \, , \vec 0 \,)$,
   where $f_\pi$ satisfies the following relation,
\begin{equation}
  \lambda f_\pi ( f_\pi^2 - v^2 ) -H=0 \,  . \label{bb}
\end{equation}
Here the parameters are chosen as $ \lambda = 20.0 , \:  v = 87.4 ~{\rm MeV} $ and
   $ H = (119 {\rm MeV} )^3 $ so that
\begin{equation}
   f_\pi = 92.5 ~{\rm MeV},   \; \; m_\pi = 135 ~ {\rm MeV},  \; \;  m_\sigma = 600 ~{\rm MeV}. \label{bc}
\end{equation}
In order to study the DCC, Rajagopal and Wilczek considered quenching in the linear sigma model.
The quenching generates an isotropic rolling down of the fields $\phi$ in isospin space from the top of the 
   Mexican hat potential.
They solved the equation of motion by a numerical simulation and found that long wavelength modes 
   of the pion fields are amplified after quenching.
The amplification is necessary for the formation of large DCC domains.
To understand the enhancement of the long wavelength modes founded by the numerical simulation, they consider
   the equation of motion for the pion modes $ {\vec \pi({\bf k},t)} $ with the approximation where the field
   $\phi^2 (x)$ is replaced by its spatial average $ \langle \phi^2 \rangle (t) $ :
\begin{equation}
 {d^2 \over dt^2}{\vec \pi({\bf k},t)}+m_{\rm eff}^2(k,t)\,{\vec \pi({\bf k},t)}=0 .
  \label{bd}
\end{equation}
The effective mass $m_{\rm eff}(k,t)$ depending on the time $t$ and the momentum $ k= \vert {\bf k} \vert $ is
   defined as
\begin{equation}
  m_{\rm eff}^2(k,t) \equiv k^2
     + \lambda \, \{ \langle \phi^2 \rangle(t)-v^2 \} .
  \label{be}
\end{equation}
The approximation, $\phi^2 (x) \rightarrow \langle \phi^2 \rangle (t) $, makes the equation of motion a linear
   equation with respect to the pion mode $ {\vec \pi({\bf k},t)} $ provided that $ \langle \phi^2 \rangle (t) $
   is given at first.
During the isotropic rolling down of $\phi$, the spatial average $ \langle \phi^2 \rangle  $ is less than $v^2$,
   so that the effective mass satisfies $m^2_{\rm eff} < 0$ for small $k^2$.
The long wavelength pion modes $k$ with $m^2_{\rm eff} < 0$ then grow exponentially.
This is the quench mechanism.

What occurs after the quench mechanism is over?
The effective mass becomes positive $m^2_{\rm eff} > 0$ after the isotropic rolling down of $\phi$.
We shall explain it in detail.
Rajagopal and Wilczek calculated the spatial average $ \langle \phi^2 \rangle (t) $ using the numerical
   solution  $\phi(x)$ of the (exact) equations of motion.
According to Fig.2 in Ref.\cite{rf:RajWil577}, $ \langle \phi^2 \rangle (t) $ oscillates around 
   $ f_\pi^2 ( \, > v^2 \, ) $
   with a frequency $ \sim 270~{\rm MeV} = 2 \,m_\pi $ 
   after the isotropic rolling down.
The spatial average is then expressed approximately as
\begin{equation}
 \langle \phi^2 \rangle (t) \approx f_\pi^2 + B^2 \cos \,(2\,m_\pi t +2\,\varphi ),
  \label{bfa}
\end{equation}
where the amplitude $B^2$ decreases with time.
After $ B^2 < f_\pi^2 - v^2 $, the spatial average $ \langle \phi^2 \rangle (t) $ is always larger than
   $v^2$, so that the effective mass becomes $m^2_{\rm eff} > 0$.
Therefore, it seems that all the pion modes are stable and oscillate after the isotropic rolling down
   $ \langle \phi^2 \rangle > v^2 $, as discussed in Ref.\cite{rf:RajWil577}.
However, the numerical simulation Fig.1 in Ref.\cite{rf:RajWil577} shows that the long wavelength pion modes
   are still amplified even when $ \langle \phi^2 \rangle > v^2 $.
Indeed, the pion mode with $ k = 0.20/a = 40 ~{\rm MeV}$ is amplified until
   $ t \sim 28 a \sim 28 \, {\rm fm/c} $, where the parameter $ a = ( 200 {\rm MeV} )^{-1} $.
How can we understand this phenomenon?

To understand the behavior of the long wavelength pion modes after the isotropic rolling down,
   we use the equation of motion  (\ref{bd}) with the same approximation,
   $\phi^2 (x) \rightarrow \langle \phi^2 \rangle(t) $.
As seen previously, the time dependent mass parameter $ m_{\rm eff}^2 (k,t) $ in Eq.(\ref{bd}) oscillates.
When the oscillating part of $ m_{\rm eff}^2 (k,t) $ has a suitable frequency, parametric resonance
   will work.
Substituting Eq. (\ref{bfa}) into Eq. (\ref{bd}), one has
\begin{equation}
{d^2 \over dt^2}{\vec \pi({\bf k},t)}+ \{(m_\pi^2 +k^2)
   +\lambda \, B^2 \cos \,(2\,m_\pi t +2\,\varphi )\, \}
  \,{\vec \pi({\bf k},t)}=0,
  \label{bg}
\end{equation}
where the term $ \lambda \, B^2 \cos \,(2\,m_\pi t +2\,\varphi ) $ is called "stimulating force"  \cite{rf:MroMul}.
Before examining this equation in detail, let us consider it qualitatively.
The eigenfrequency $ \omega_\pi $ of the long wavelength modes of the pion is about $ m_\pi \, $,
   $ \omega_\pi \equiv \sqrt{m_\pi^2 + k^2} \approx m_\pi $ , and the frequency $ \omega_{\rm s} $
of the stimulating force is $ 2 \, m_\pi \,$,  $ \omega_{\rm s} = 2 \, m_\pi $. Then, the parametric resonance
works for the long wavelength modes, because
   $ \omega_{\rm s} \approx 2 \, \omega_\pi \,$
    \cite{rf:LanLif}.
Consequently, the long wavelength modes of the pion with $ k \ll m_\pi $ are amplified by the parametric
   resonance after the quench mechanism is over.

Here, our equation (\ref{bg}) should be compared with that of Mr\'owczy\'nski and M\"uller \cite{rf:MroMul}
   to clarify the difference between them.
They assume that, after the rolling down, the field $\phi$ oscillates only along the sigma direction.
Since the parametric resonance induced by the $\sigma$ field oscillation is concentrated, the equation for the
   pion modes $ {\vec \pi}^{(1)}_{\rm MM}({\bf k},t) $ has the form \cite{rf:MroMul,rf:HirMin},
\begin{equation}
{d^2 \over dt^2}{\vec \pi}^{(1)}_{\rm MM}({\bf k},t)+ \{(m_\pi^2 +k^2)
   -2\lambda v \sigma_0 \cos \,(m_\sigma t + \varphi )\, \}
  \,{\vec \pi}^{(1)}_{\rm MM}({\bf k},t)=0 ,
  \label{bh}
\end{equation}
where the frequency $ \omega'_{\rm s} $ of the stimulating force is $ m_\sigma \,$,
    $ \omega'_{\rm s} = m_\sigma $.
In this theory  (\ref{bh}), if the sigma's amplitude $ \sigma_0 $ is not large,
   the long wavelength modes of the pion with $ k \ll m_\pi $ are not amplified by the
   parametric resonance  because $ \omega'_{\rm s} \gg 2 \, \omega_\pi \,$.
On the other hand, in our theory (\ref{bg}), the long wavelength modes $ k \ll m_\pi $ are amplified
   because of $ \omega_{\rm s} \approx 2 \, \omega_\pi \,$, as seen above.

Let us go back to the discussion of our theory.
To analyze Eq. (\ref{bg}) more quantitatively, we introduce a new variable,
   $ z \equiv \,m_\pi \, t +\,\varphi \,$.
The equation (\ref{bg}) is then rewritten as
\begin{equation}
{d^2 \over dz^2}{\vec \pi({\bf k},z)}+ \{A
   + 2 q \cos \,(2z )\, \}
  \,{\vec \pi({\bf k},z)}=0 ,
  \label{bi}
\end{equation}
where
\begin{equation}
  A \equiv {m_\pi^2 + k^2 \over m_\pi^2}~ , \hskip2cm
   q \equiv {\lambda B^2 \over 2 \, m_\pi^2 }~,
  \label{bj}
\end{equation}
and $B^2$ decreases with time.
For simplicity, we treat Eq.(\ref{bi}) at short intervals of time, in which $B^2$ can be considered
   approximately as a constant.
Eq.(\ref{bi}) is called the Mathieu equation \cite{rf:MroMul,rf:LanLif}, whose unstable solution increases
   rapidly with time.
The unstable solutions  $ {\vec \pi}( {\bf k} , z ) $ of the  Mathieu equation in the first resonance satisfy the
   following condition $ \vert A -1 \vert < q + O(q^2 ) \,$ if $ \, 0<A <2 \, $ and $ q <1 $.
Therefore, the long wavelength modes with
\begin{equation}
  k \, < \, B \sqrt{\lambda \over 2} \, ,
  \label{bk}
\end{equation}
are amplified by the parametric resonance.

Does the condition  (\ref{bk}) explain quantitatively the enhancement of the long wavelength pion modes
   observed in the numerical simulation of Ref.\cite{rf:RajWil577} ?
We shall focus on two stages of the parametric resonance: (i) at the early stage and (ii) at the late stage.
(i) At the early stage with the amplitude $  B^2=f_\pi^2 - v^2 = m_\pi^2 /  \lambda \sim ( 30 {\rm MeV} )^2 $
    corresponding to the time ~$ t \sim 10 a \sim  10 \, {\rm fm/c} $ in Fig.2 of Ref.\cite{rf:RajWil577},
    the space average $ \langle \phi^2 \rangle (t) $ oscillates as 
   $ v^2 \leq \langle \phi^2 \rangle (t) \leq  2 f_\pi^2 - v^2 $.
The condition  (\ref{bk}) predicts that the pion modes with
\begin{equation}
  k < {m_\pi \over \sqrt{2} } = 96 ~ {\rm MeV},
  \label{bl}
\end{equation}
are amplified when ~$ t \sim  10 \, {\rm fm/c} $.
This prediction is verified if one compare it with the result of the numerical simulation Fig.1 of
   Ref.\cite{rf:RajWil577}.
(ii) At the late stage with the amplitude $  B^2 \sim ( 15 {\rm MeV} )^2 $
   corresponding to the time
   ~$ t \sim 26 \, {\rm fm/c} $ in Fig.2 of Ref.\cite{rf:RajWil577},
   $ \langle \phi^2 \rangle (t) $ oscillates as
   $ v^2 < \langle \phi^2 \rangle (t) <  2 f_\pi^2 - v^2 $.
This time the condition  (\ref{bk}) predicts
\begin{equation}
  k < 47 ~ {\rm MeV}\, .
  \label{bm}
\end{equation}
This prediction is also verified in the same manner.
Thus the parametric resonance  (\ref{bk}) is able to explain quantitatively the enhancement of the long 
   wavelength modes of the pion $ 0 < k < m_\pi $ observed in the numerical simulation \cite{rf:RajWil577}.
The characteristic features of the enhancement after the quench mechanism are (a) that the long wavelength
   modes are amplified for more than $ 10~{\rm fm/c} $ and (b) that the lower the momentum $k$ of the pion is,
   the longer a span of the enhancement becomes.
The time evolution of the zero mode $k=0$  of the pion will be discussed in the next section.
%
%
%
%
%
\section{Pion's zero mode oscillation}
In the parametric resonance mechanism, the oscillating part of $ \langle \phi^2 \rangle (t) $ plays a role
   of the "stimulating force".
What is the origin of the oscillation part $ B^2 \cos \,(2\,m_\pi t +2\,\varphi ) $ having the frequency
   $2 \, m_\pi $ ?
The spatial average $ \langle \phi^2 \rangle (t) $ is written by the sigma modes
   $ \sigma ( {\bf k}, t ) $ and the pion modes $ \pi_a ({\bf k}, t ) $ as
\begin{equation}
  \langle \phi^2 \rangle (t) = 
   \int {d{\bf k} \over (2\pi)^3} \, \vert \sigma ( {\bf k} , t ) \vert^2
   + \sum_{a=1}^{3} \int {d{\bf k} \over (2\pi)^3} \, \vert  \pi_a ( {\bf k} , t ) \vert^2 \, .
  \label{ca}
\end{equation}
According to the numerical simulation in Fig.1 of Ref.\cite{rf:RajWil577}, the frequency of
   $ \vert \pi_a ( {\bf k}, t )\vert ^2 $ is given by $ 2 \sqrt{m_\pi^2 + k^2}$, while the frequency of
   $ \vert \sigma ( {\bf k}, t ) \vert ^2 $ can not be specified due to its complicated behavior.
Therefore, the $ \langle \phi^2 \rangle $'s oscillating part $ B^2 \cos \,(2\,m_\pi t +2\,\varphi ) $
   arises from the zero mode $k=0$ of the pion.
   
Now, we suppose that the oscillation of the pion's zero mode is represented as the following simple form
\begin{equation}
    \langle \pi_a \rangle (t) = \gamma_a \cos \, ( m_\pi t + \varphi ) ,
      \hskip1cm  (a=1,2,3 ) ,
  \label{cb}
\end{equation}
in order to find the relation between  $B^2$ and the amplitude $\gamma_a$ of the pion's
   zero mode oscillation.
We obtain
\begin{equation}
   B^2 \cos \,(2\,m_\pi t +2\,\varphi )
   = {1 \over 2} \sum_{a=1}^{3} \gamma_a^2   \cos \, ( 2m_\pi t + 2\varphi ) ,
  \label{cc}
\end{equation}
because the origin of the oscillating part of $ \langle \phi^2 \rangle (t) $ is the zero mode of the pion.
Consequently, it yields a relation
\begin{equation}
  B^2 = {1 \over 2}  \sum_a \gamma_a^2        \: .
  \label{cd}
\end{equation}
For example, if the zero mode has a form
   $ \langle {\vec \pi} \rangle(t) = ( \gamma, \gamma, \gamma/4 ) \cos \,(m_\pi t+ \varphi ) $, the relation
    (\ref{cd}) becomes $B = \sqrt{33/32} \, \gamma$.
Let us estimate the pion's zero mode amplitude $\gamma$ at the two stages of the parametric resonance
   (i) and (ii) discussed in the previous section.
(i) At the early stage ( $ t \sim 10 \, {\rm fm/c}, \,   B^2 = f_\pi^2 - v^2  $ ), the magnitude of the amplitude
   is $\gamma \sim 0.32 \, f_\pi $.
(ii) At the late stage ( $ t \sim 26 \,{\rm fm/c}, \,   B^2 = ( 15 {\rm MeV} )^2   $ ), 
   that is $\gamma \sim 0.16 \, f_\pi $.
   
According to the numerical simulation Fig.2 of Ref.\cite{rf:RajWil577}, the amplitude $B^2$ decreases with time.
The relation  (\ref{cd}) implies that the amplitude $\gamma_a$ of the pion's zero mode also decreases.
This leads us to the following problem.
The fact $\gamma_a$ decreases with time contradicts with the statement in the previous section: the statement
   is that the zero mode $k=0$ of the pion field belongs to the unstable solutions  (\ref{bk}) of Eq.(\ref{bg}),
   so that the zero mode is amplified.
This contradiction is, however, resolved since the approximation used in Eq.(\ref{bg}) is not applicable
   to the zero mode.
The reason is as follows.
The equation of motion (\ref{bg}) with the approximation is regarded as a linear equation with
   respect to the pion mode $ \pi_a ( {\bf k}, t ) $.
Hence the nonlinear terms of $ \pi_a ( {\bf k}, t ) $ involved in the exact equation of motion are not concerned.
But, in the equation of motion for the zero mode $k=0$, the effect of nonlinear terms of
   zero mode is not negligible \cite{rf:Kai}.
%
%
%
%
%
\section{Summary and discussion}
We studied the classical linear sigma model to investigate the disoriented chiral condensate ( DCC ).
We showed that the long wavelength modes of the pion are amplified for more than $ 10~{\rm fm/c} $ by
   the parametric resonance.
The parametric resonance mechanism follows the quench mechanism, both of which enhance 
   the long wavelength modes of the pion.
The shifting from the quench mechanism to the parametric resonance mechanism is described by a time dependent
   quantity.
That is the spatial average $ \langle \phi^2 \rangle (t) $:
\begin{eqnarray}
    \langle \phi^2 \rangle (t)
     \left\{  \begin{array}{ll}
           < v^2  \, ,                                  &  \;  \mbox{    early time }    \\
          \approx f_\pi^2 + B^2 \cos \, ( 2 \, m_\pi t + 2 \, \varphi )  \, ,  &  \;  \mbox{  late time .  } 
                 \end{array}
     \right.
 \label{da}
\end{eqnarray}
During $ \langle \phi^2 \rangle (t) < v^2 $, the quench mechanism works as argued by Rajagopal and Wilczek.
After that, $ \langle \phi^2 \rangle (t) $ oscillates around $f_\pi^2$ with the frequency $ 2 \, m_\pi $.
This oscillation plays a role of "stimulating force" for the pion modes $ k ( > 0 ) $.
Since the frequency of the stimulating force is about twice as much as the eigenfrequency of the long wavelength
   pion modes, the parametric resonance mechanism works.
That mechanism can explain quantitatively the result of numerical simulation in Ref.\cite{rf:RajWil577} 
   concerning the amplification of the long wavelength modes of the pion.
   
As stated in the previous section, the "stimulating force" with the frequency $ 2 \, m_\pi $ is attributable to
the
   zero mode of the pion.
Therefore, we believe that the oscillation of the pion's zero mode $k=0$ around the true vacuum causes the 
   parametric resonance by which the long wavelength modes  $k \,  ( > 0 ) $ of the pion are amplified.
This scenario is contrast with the scenario of Refs.\cite{rf:MroMul,rf:HirMin,rf:Kai} in which 
   the oscillation of the sigma's zero mode causes the  parametric resonance.
%
%
%
%
%
%
%
\newpage
\end{document}